\documentclass[a4paper]{jpconf}

\usepackage{color}  

\usepackage{graphicx}
\usepackage{amsmath,amssymb}
\usepackage{latexsym, ulem,bm}

\begin{document}
\title{Conformal amplitude hierarchy and the Poincar\'{e} disk}
\author{Hirohiko Shimada}
\address{Mathematical and Theoretical Physics Unit, Okinawa Institute of Science and Technology Graduate University, Onna, Okinawa, 904-0495, Japan}
\ead{hirohiko.shimada@oist.jp}

\begin{abstract} 
The amplitude for the singlet channels in the 4-point function of the fundamental field in the conformal field theory of the 2d $O(n)$ model is studied as a function of $n$. 
For a generic value of $n$, the 4-point function has infinitely many amplitudes, whose landscape can be very spiky as the higher amplitude changes its sign many times at the simple poles, which generalize the unique pole of the energy operator amplitude at $n=0$.
In the stadard parameterization of $n$ by angle in unit of $\pi$, we find that
the zeros and poles happen at the rational angles, forming a hierarchical tree structure inherent in the Poincar\'{e} disk. Some relation between the amplitude and the Farey path, a piecewise geodesic that visits these zeros and poles, is suggested.
In this hierarchy, the symmetry of the congruence subgroup $\Gamma(2)$ of $SL(2,\mathbb{Z})$ naturally 
arises from 
the two clearly distinct even/odd classes of the rational angles, in which one respectively
gets   
the truncated operator algebras and the logarithmic 4-point functions.
\end{abstract}

\section{Introduction} 
\vspace{3mm}
\paragraph{\underline{The amplitudes in CFT$_d$}}
In general, a CFT  has 
the 3-point functions $\langle\phi_i\phi_j\phi_k\rangle$ of the fields with 
the scaling dimensions $(\Delta_{i}, \Delta_{j}, \Delta_{k})$
taking the power law form 
$\langle\phi_i(x_1)\phi_j(x_2)\phi_k(x_3) \rangle$$=$$C_{ijk}$   
$|x_{12}^2|^{\frac{\Delta_{i}+\Delta_{j}-\Delta_{k}}{2}}$ 
$|x_{23}^2|^{\frac{\Delta_{j}+\Delta_{k}-\Delta_{i}}{2}} $
$|x_{31}^2|^{\frac{\Delta_{k}+\Delta_{i}-\Delta_{j}}{2}}$
 with a set of unknown numbers
$C_{ijk}$ and the 4-point function $\langle\phi\phi\phi\phi\rangle$ of the same field 
with the scaling dimension $\Delta_\phi$
taking the form 
$\langle\phi(x_1)\phi(x_2)\phi(x_3)\phi(x_4) \rangle$
$=g(u,v) (x_{12}^2x_{34}^2)^{-\Delta_\phi}$  
with an unknown function $g(u,v)$
of the cross ratios $u=\frac{x_{12}^2x_{34}^2}{x_{13}^2x_{24}^2}$ 
and $v=\frac{x_{14}^2x_{23}^2}{x_{13}^2x_{24}^2}$ \cite{polyakov}. 
Although the function $g(u,v)=g_{\phi}(u,v)$ depends on the choice of the field $\phi$ via the possible selection rule, hereafter the subscript $\phi$ is omitted in the simplest setting, where we are mostly interested in the 4-point function of the fundamental field.

On top of these kinematical constraints, the dynamical constraints come from the associativity of the operator algebra, which can be formulated as the crossing symmetry of the 4-point function 
\begin{align}
v^{\Delta_\phi} g(u,v) = u^{\Delta_\phi} g(v,u)
\label{crossing}
\end{align}
The function $g(u,v)$ starts from $ 1 $ in the factorization limit 
$\langle\phi\phi\phi\phi\rangle\to \langle\phi\phi\rangle\langle\phi\phi\rangle$, and the sub-leading terms are a sum over the contribution from the intermediate channel labeled by $ k $:
\begin{align}
g(u,v) = 1+\sum_{k=1}^{\infty}\left[C_{\phi\phi}^k\right]^2 G_{k}(u,v)
\end{align}
Each term is a conformal block $G_k$ which comes with some amplitude in the form of $C^2$ as a result of applying twice the operator product expansions (OPE) \footnote{
The OPE coefficient $C_{\phi\phi}^k$ is the same thing as $C_{ijk}$ in the 3-point function of $\phi_i=\phi, \phi_j=\phi$, and $\phi_k$ under the natural normalization of the 2-point functions $\langle\phi(x_1)\phi(x_2) \rangle=(x_{12}^2)^{-\Delta_\phi}$
}. Then the crossing symmetry \eqref{crossing} gives highly non-trivial constraints on the amplitudes $\left[C_{\phi\phi}^k\right]^2$.
Only later in Section \ref{section:odd},
we will need the conformal block $G_k$. For the time being, we focus on 
the amplitude.

\paragraph{\underline{The positivity of the amplitudes}}
Another interesting extra assumption is the unitarity, which requires all the amplitudes be positive:
 $\forall k~~ \left[C_{\phi\phi}^k\right]^2 >0$.
Indeed, the nice developments comes from this unitarity restriction: By using the linear functionals \cite{ising1,kos1}, 
it is possible to judge whether any {\it unitary} solutions to \eqref{crossing} can exist for a given
set of the lowest scaling dimensions, namely, that of the spin operator and the energy operator.
With some reflection on the low-lying spectrum, one certainly expect some boundary curve will appear. In retrospect, this is not surprising. 
What is surprising is that this bound is not monotone, but it has a pronounced kink, 
which turns out to be located at the scaling dimensions where the 3D Ising model has to be \cite{ising1, ising2}.  
We have yet to understand the real background of this phenomenon.    

If we try to avoid being tied down to the kink, and instead, try to generalize the problem as far as possible, 
one would recognize that the perfect unitarity is very special, isolated phenomenon.
Unitarity is always violated except the case where both $d$ and $n$ are integers. 
In $d=4-\epsilon$ dimensions \cite{hogervorst},
it is suggested that the unitarity is violated very weakly in the sense that the negative amplitude occurs only at the operators with high scaling dimensions, and if it occurs, with very small amplitudes.
Such scenario is consistent with the observation that one could locate the 3d $O(n)$ CFTs rather successfully until it gets really close to $n = 0$ \cite{shimada}.  Here we will discuss (1) the unitarity 
violation occurs at $n=0$ at the main singlet amplitude  (i.e. the energy operator), and (2) it is the root of the other relatively minor violations.  

This note is intended for giving crude ideas and tangible examples. 
 In Section 2, after some warm up, the computation of certain amplitudes in the 2d 4-point function $\langle\sigma\sigma\sigma\sigma\rangle$ is outlined omitting the details. 
In Section 3, all the location of the zeros and poles are identified: the Stern-Brocot tree structure and
the $\Gamma(2)$ symmetry \cite{simon} of a hyperbolic space arise naturally. 
In Section 4, we see there are clearly distinct even/odd classes of the 4-point functions. 
The even one is rather trivial, and is related to the finite truncation of the OPE, while
the odd one has more interesting structure leaving the logarithmic contribution to the 4-point functions.

\section{Amplitudes as functions on $(d, n)\in \mathbb{R}^2$}


\subsection{a global view of the universality class}
The main body of this note deals with an interesting analytic structure of the conformal amplitude in $d=2$. Yet, in order not to overlook the important physics underlying the conformal amplitude, it is perhaps good to start with the Ising model, with which most readers are familiar with. It has the Hamiltonian
\begin{align}
H= J \sum_{\langle ij\rangle} s_i \cdot s_j
\end{align}
 where $s_i$ is the spin variable on site $i$ taking the value $+1$ or $-1$, and $\langle ij\rangle$ stands for the nearest-neighbor sites on a $d$-dimensional lattice.  
At the critical point this model shows the scale invariant pattern, and the continuum Hamiltonian with a double-well potential ($\varphi^4$ model)  well describes the vicinity of the critical point.
A slight generalization of the internal symmetry: $\mathbb{Z}^2\to O(n)$ yields a Hamiltonian
\begin{align}
H=\int d^d x~  (\partial \vec{\varphi})^2 - m^2 \vec{\varphi}^2 + \lambda (\vec{\varphi}^2)^2,
\label{phi4}
\end{align} 
where the fundamental field is the $n$-component vector $\vec{\varphi}=\phi_{a}$ $(a=1,\cdots \, n)$. 
The model at $n=1$ corresponds to
 the Ising model, and the interacting CFT is realized as the IR fixed point of the renormalization group.

Now a particularly important idea is to regard the space dimension $d$ as a continuous variable \cite{wilson}.
This led us to the idea of non-trivial fixed points below $d=4$, which passes through $d=2$ as the 
well-understood $c=1/2$ CFT (Majorana fermion). 
On the other hand, a detailed analysis of 3d CFTs is still challenging; in the long run, it would be nice if one could somehow extend the analytic techniques in 2d to 3d. 

Consider the global view of the CFTs on $(d, n)$-plane.
Here it is very important to regard also $n$ as a continuous parameter.
For instance, the $n\to 0$ limit describes the physics of the polymer in a good solvent \cite{degennes}.
This problem is revisited \cite{shimada} by using the modern technique 
\footnote{An up-to-date table of the state-of-art results by various methods may be found in \cite{schram}.}
of the conformal bootstrap to obtain the estimate for the fractal dimensions $d_F=1.7016(36)$.

In 2d, beyond $n=2$ the target space $S^{n-1}=O(n)/O(n-1)$ becomes essentially non-abelian, resulting in a massive theory.
The one parameter family of CFT \cite{difrancesco,jacobsen} is studied for $|n|\leqslant2$. 
Whole family can be  studied in the Schramm Loewner Evolution \cite{bauer}, which is an attempt at describing various critical fractals by using the conformal map driven by the stochastic Brownian motion of strength $\kappa$.
There is also nice formula for the fractal dimension proved by Beffara \cite{beffara}. 
Along with the central charge, they are given by
\begin{align}
d_F=1 + \frac{\kappa}{8} ,\quad n=-2\cos 4\pi/\kappa, \quad \kappa/4=\rho=1/g, \quad 2\leqslant \kappa \leqslant 4
,\quad c=1-6(g-1)^2/g,
\label{formulaBeffara}
\end{align}
where we have also indicated the relation between $\kappa$ and the Coulomb gas couplings $g=\alpha_+^2$, $\rho=\alpha_-^2$. In Section \ref{section:poincare}, the coupling $g$ measures the angle along the boundary of the Poincar\'{e} disk. 


\subsection{
The  4-point function of the fundamental fields $\langle\sigma\sigma\sigma\sigma\rangle$ is non-degenerate} 
Let us then turn to our current focus, the 4-point function in $d=2$.
BPZ shows that a 4-point function satisfies a differential equation of order $r\cdot s$ \cite{belavin} if it contains at least one {\it degenerate} field 
$\phi_{r,s}$ labeled by a pair of positive integers $(r,s)$ (the Kac labels).
But unfortunately, we do not know how to compute the 4-point function of the fundamental fields, that lies at the heart of the bootstrap program. 
This is because, according to the spectrum obtained from the torus partition function \cite{difrancesco}, the spin field is identified with $(r,s)=(1/2,0)$: $\sigma\to\phi_{\frac{1}{2},0}$, where the label is not a pair of positive integers. So they are non-degenerate.

On the other hand the energy field is identified with $(r,s)=(1,3)$: $\varepsilon\to \phi_{1,3}$. 
They are degenerate, and satisfies $3$-rd order differential equation. 
Thus, we also know $\langle\sigma\sigma\varepsilon\varepsilon\rangle$ has exactly three channels \cite{dotsenkofateev}.
In contrast, $\langle\sigma\sigma\sigma\sigma\rangle$ may have infinitely many intermediate states in general.
Such cases is more difficult, but they deserve more attention as the interacting CFT in $d>2$ has typically infinitely many channels. 




\subsection{Double factorization of the integral}
The main issue here is that there seems to be no simple way 
\footnote{As a technical remark, if $(2r,2s)$ were a pair of {\it positive} integers such as the case with $(r,s)=(1/2,3/2)$, one could still devise an integral formula by introducing the integer numbers 
of the screening charges $\alpha_\pm$. 
Such cases had been known to be important in the geometrical lattice models 
(the polymer CFT \cite{saleur} for example) 
and has been revisited for the percolation CFT \cite{dotsenko}. 
}
of working out the fusion rule between the non-degenerate fields 
$\phi_{r, s}$ with $(r,s)=(1/2, 0)$.  Still one should try to infer the fusion rule indirectly.
 When the energy operator $\varepsilon=\phi_{1,3}$ is fused by itself,  it creates  
a new higher operator $\phi_{1,5}$ as well as the identity operator $\phi_{1,1}$ and the energy operator itself since
$\phi_{1,3}\times \phi_{1,3}=\phi_{1,1}+\phi_{1,3}+\phi_{1,5}$.   
By repeated fusions, it is natural to think that the algebra in the $O(n)$ CFT for generic $n\in \mathbb{R}$
may contain $\varepsilon^{(p)}=\phi_{1, 2p+1}$ $(p=0,1,\cdots)$. 
In terms of the fields in the Hamiltonian \eqref{phi4}, the correspondence should look like $\varepsilon^{(p)}=:(\vec{\varphi}^2)^p:$. 
By symmetry, it is natural to have
\begin{align}
\sigma\times \sigma = \varepsilon^{(0)} + \varepsilon^{(1)} + \varepsilon^{(2)}+ \varepsilon^{(3)} +\cdots
\label{sigmasigma}
\end{align}
Note that this expression focus on the $O(n)$ {\it singlet} channels, 
and the contribution of the other types, if exists, may be omiteed.

We take the following strategy used in Dotsenko-Fateev: instead of computing 
$\langle\sigma\sigma\sigma\sigma\rangle$,
we first compute the amplitudes $B_k$ in the mixed correlator 
$\langle\sigma\sigma\varepsilon^{(p)}\varepsilon^{(p)}\rangle$, which contains the both the 
{\it non-degenerate} field $\sigma$ and 
the degenerate field $\varepsilon^{(p)}$, 
and the other amplitudes $A_k$ in the 4-point function $\langle\sigma\sigma\varepsilon^{(p)}\varepsilon^{(p)}\rangle$ of the 
degenerate field $\varepsilon^{(p)}$ only. 
Concretely, these two correlation function reads
\begin{align}
\langle\sigma(0,0)\sigma(z,\bar{z})\varepsilon^{(p)}(1,1)\varepsilon^{(p)}(\infty)\rangle= |z|^{2\Delta_\sigma}|z-1|^{2\Delta_{\varepsilon^{(p)}}}
&\sum_{k=0}^N B_k~|z|^{2\lambda_k^{(B)}}
f_k^{(B)}(z, \bar{z}),\\
\langle \varepsilon^{(p)}(0,0)\varepsilon^{(p)}(z,\bar{z})\varepsilon^{(p)}(1,1)\varepsilon^{(p)}(\infty)\rangle=|z(z-1)|^{2\Delta_{\varepsilon^{(p)}}}
&\sum_{k=0}^N A_k~|z|^{2\lambda_k^{(A)}}
f_k^{(A)}(z, \bar{z}),
\label{AkBk}
\end{align}
where $|z|^{2\lambda_k}f_k(z,\bar{z})=|\mathcal{F}(c, \lambda_k, \{\Delta_i\} | z)|^2$ is a product of two chiral conformal blocks. 
The function $f_k(z,\bar{z})=1+\mathcal{D}(z,\bar{z})$  is a universal object, with which the form of the descendant 
contribution $\mathcal{D}(z,\bar{z})$ is entirely determined from the conformal invariance as an infinite series in $z$ and $\bar{z}$, and carries no dynamical information.
The exponent $\lambda_k$ is the scaling dimension  $\Delta_{\varepsilon^{(k)}}$, which plays no role until
Section \ref{section:odd}.
The main objects here are $A_k$ and $B_k$, which are products of two OPE coefficients:
$A_k=\left[C^{\varepsilon^{(p)}}_{\sigma\sigma}\right]^2$ and
$B_k=\left[C^{\varepsilon^{(p)}}_{\sigma\sigma} C^{\varepsilon^{(p)}}_{\varepsilon^{(p)}\varepsilon^{(p)}} \right]$.
Then one may take the ratio and eliminates the unwanted OPE coefficient $C^{\varepsilon^{(p)}}_{\varepsilon^{(p)}\varepsilon^{(p)}} $
, yielding the amplitude of our interest
\begin{align}\label{ratio}
\left[C^{\varepsilon^{(p)}}_{\sigma\sigma}\right]^2=B^2_k/A_k.
\end{align}
The normalization to $\left[C^{\varepsilon^{(0)}}_{\sigma\sigma}\right]^2=1$ can be done
by dividing this by $B_0^2/A_0$.
Both $A_k$  and $B_k$ can be computed as the same coefficient $Q^{(N)}_k$ in the expansion of the following symmetric complex integral: 
\begin{align}\label{IN}
I_N(a,b,c,\rho; z, \bar{z})&=\int 
\prod_{j=0}^{N-1} \left[d^2t_j |t_j|^{2a}
|1-t_j|^{2b} |t_j-z|^{2c}
\prod_{i(\neq j)}^{N-1}
|t_i-t_j|^{2\rho}\right]\\
&=\sum_{k=0}^{N} Q^{(N)}_k |z|^{2\lambda_k}\left( f_k(z,\bar{z}) \right),
\end{align}
where, though the detail of the vertex operator representation is omitted,
the exponents satisfy\footnote{
If one adopts another known identification \cite{dotsenkofateev}
$\sigma\to \phi_{\frac{m-1}{2},\frac{m+1}{2}}$ with $\rho=\frac{m}{m+1}$, 
the exponents $a$ and $c$ are interchanged, giving the same amplitudes.
Indeed, these two fields are related by the symmetry \eqref{secret}.
}
 $a+c=2\cdot 2\alpha_0\alpha_-=2\rho-2$ and $b=2\alpha_{1,N+1}\alpha_-=-N\rho$
with $N=2p$.
The Coulomb gas screening charges $\alpha_{+}>0$ and $\alpha_{-}<0$ 
are given in \eqref{formulaBeffara}. 

The first step in taming such integrals
often involves solving the monodromy constraints for the bilinear form of the chiral integrals \cite{dotsenko,estienne}.
We find it simpler to take the following heuristic approach. One first chooses $k$ out of $N$ variables in \eqref{IN}
 and scales them by the cross ratio: $t_j \to  z t_j$.
The result is the factorization of $Q^{(N)}_k$ into the two integrals  
\begin{align}\label{key}
Q^{(N)}_k= \frac{N!}{k!(N-k)!} \mathcal{C}_{k}(1+a,1+c,\rho)\mathcal{C}_{N-k}(1+a+c+2k\rho, 1+b, \rho). 
\end{align}
The complex Selberg integral $\mathcal{C}_K$ is defined as
\begin{align}
\mathcal{C}_K(\alpha,\beta,\gamma)=\int d^2t_1\cdots d^2t_K 
\prod_{j=0}^{K-1} |t_j|^{2(\alpha-1)}
|1-t_j|^{2(\beta-1)} \cdot
\prod_{0\leqslant i<j<K}
|t_i-t_j|^{4\gamma}\,
,
\end{align}
which, in turn, factorizes into the square of the (usual) chiral Selberg integral 
\cite{selberg}
times an interesting trigonometric factor \cite{aomoto}
\begin{align}
\mathcal{C}_K(\alpha,\beta,\gamma)&=S_K^2(\alpha,\beta,\gamma) \,
\frac{1}{K!} \prod_{j=0}^{K-1} \frac{\sin \pi(\alpha+j\gamma)
\sin\pi(\beta+j\gamma) \sin\pi (j+1)\gamma}
{\sin \pi (\alpha+\beta+(K+j-1)\gamma) \sin\pi\gamma},\\
S_K(\alpha,\beta,\gamma) & =
\int_0^1 \cdots \int_0^1 \,
 \prod_{j=1}^{K-1} t_i^{\alpha-1}(1-t_i)^{\beta-1}
\prod_{0\leqslant i < j< K} |t_i - t_j|^{2\gamma}\,
d t_0\cdots d t_{K-1} \nonumber\\
&=
\prod_{j=0}^{K-1} \frac{\Gamma (\alpha+j\gamma)
\Gamma(\beta+j\gamma)\Gamma(1+(j+1)\gamma)}
{\Gamma(\alpha+\beta+(K+j-1)\gamma)\Gamma(1+\gamma)}.
\end{align}
This factor guarantees the monodromy invariance of the 4-point function solved for the simplest case 
by Dotsenko-Fateev. The monodromy constraints was also solved in a useful form \cite{estienne}.
In contrast, the integral formula \eqref{IN} is based on the full vertex operator, and then the monodromy invariance follows automatically.
The amplitude in $\langle\sigma\sigma\sigma\sigma\rangle$ yields
\begin{align}
\left[C^{\varepsilon^{(p)}}_{\sigma\sigma} \right]^2=\frac{\mathcal{Z}(\rho)}{\gamma((2p+1)\rho)}\prod_{j=1}^{p}\frac{\gamma^2(j \rho)}{\gamma^2((p+j)\rho)} ~\uline{\gamma^2\left(j \rho -\frac{1}{2} \right)\gamma^2\left(j \rho +\frac{1}{2} \right)},
\label{CpssE}
\end{align}
with the less important factor $\mathcal{Z}(\rho)$, 
which has no singularities in $\frac{1}{2}<\rho<1$ ($|n|<2$).
The function $\gamma(x)=\Gamma(x)/\Gamma(1-x)$ has zeros at positive integers. The factor indicated by the underline is unique
\footnote{Compared with a generic 3-point function \cite{ikhlef,dotsenko3} in the
$c\leq 1$ Liouville theory, the special case \eqref{CpssE} has unique properties, such as 
the even/odd distinction phenomenon in Section \ref{evenodd}.}
to the 4-point function of the fundamental field $\sigma=\phi_{\frac{1}{2},0}$. 

For the $O(n)$ CFT to be unitary, this amplitude should be positive for all $p\in \mathbb{N}$.
In $d=2$, it had been known \cite{difrancesco} that the perfect positivity is realized only at $n=1$ (Ising model) or at $n=2$ (XY model). 
As they are well-understood, we are more interested in generic $n$ and study how the positivity is violated.
The formula \eqref{CpssE} should capture the primary part of the positivity condition
used in the conformal bootstrap in the simplest setting using $\langle\sigma\sigma\sigma\sigma\rangle$.  Understanding the sign changes of \eqref{CpssE} is therefore important.
Identifying the location of all the singularities in the next section is an essential step to achieve this.





\section{An anatomy of the conformal amplitude}
\subsection{General structure: the spiky landscape with fast decay as the depth $p$ increases}
In Figure \ref{figure:poincare}-(b), the amplitudes \eqref{CpssE} is plotted in log-scale, in which one sees many spikes.
These spikes are either zeros or poles, and the dashed curve represents negative amplitudes.
The horizontal axis is the number of components $n$ in the parameter $\rho$ as in \eqref{formulaBeffara}.
This amplitude decays very fast as $p$ increases. For instance, $p=4$ for generic $n>0$ is already the scale of the inverse Avogadro number $\sim 10^{-23}$, thus the amplitudes with $p>4$ are literally microscopic for statistical physics application. Although the decay of the absolute value is also of interest to quantify the degree of the unitarity violation \cite{hogervorst, shimada}, 
in the following,
we focus our attention to
the sign behavior of the amplitude  \eqref{CpssE}. 

\begin{figure}
\begin{center}
\includegraphics[width=14cm]{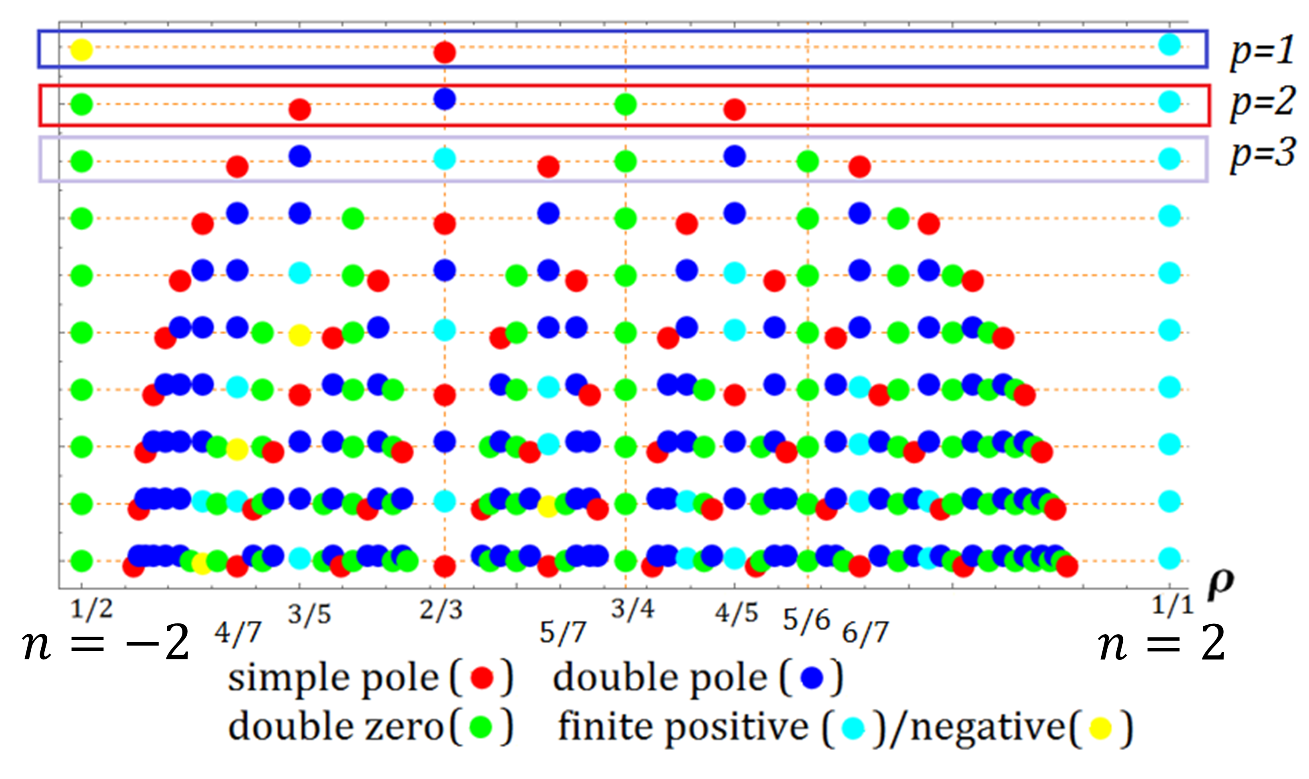}
\vspace{-0.8cm}
\end{center}
\caption{\label{figure:Farey} 
Locations of all the zeros and poles for the amplitude with the level $p\leqslant 10$.  
At given level $p$, they may be found when the parameter $\rho$ belongs to the Farey sequence $F_{2p+1}$. There are no simple zeros. The zeros of even order (green) lie at the $\rho$ with even denominators while simple (red)  and double (blue) poles lie at $\rho$ with odd denominators.
In $F_{2p+1}$, only limited points are finite whose sign can be either positive (light-blue) or negative (yellow). 
Slight vertical shifts of the markers are introduced just for avoiding the overlaps. 
 }
\end{figure}

\subsection{Stern-Brocot tree and the even/odd denominator classes}\label{section:stern}
With more detailed inspection, one would notice that the location of the singularities 
in the $\rho$-coordinates may be organized in a tree structure, namely, the Stern-Brocot tree.
In this tree, the next level is generated by the mediant 
rule\footnote{Or the Farey addition rule, see a nice exposition \cite{mumford} for the mathematical background
beyond puzzle games. } 
\begin{align}
\frac{b}{a}~\dot{+}~ \frac{d}{c} = \frac{b+d}{a+c},
\end{align}
where one sums respectively the numerators and denominators.
Then each row consists of irreducible fractions with denominator less than a certain integer, namely, the Farey sequences.  
The locations of singularities are summarized in the diagram in Figure \ref{figure:Farey}.
The amplitude at level $p$ in \eqref{CpssE} has the zeros and poles on the Farey sequence $F_{2p+1}$:
the set of irreducible fractions $\rho=m/m'$ with $m'\leqslant 2p+1$.
The point of this note is considering both the horizontal (deforming the CFT continuously)
and the vertical (going higher levels of operator in a given CFT) directions of this diagram.

Horizontally, one finds no zeros of odd order including simple zeros.  Thus the unitarity violation is triggered not by the zeros, but by the poles of the amplitude. 
At the $n=2$ (XY model), 
since the CFT is unitary, all the amplitudes for $p\in \mathbb{N}$ are positive, which may have relation with the multiple-vortex spectrum. 
The amplitude for the energy operator ($p=1$) changes its sign at the unique simple pole at $n=0$
 indicated by the red dot at $\rho=2/3$.
The positivity violation starts from here, and the others for higher $p$ may be considered as the generalization of it. 
The amplitude at some limited members of the Farey fraction is finite,
whose sign is determined by the number of the simple poles in the right of it.

Vertically, for an even denominator $\rho$, we see a series of the infinite zeros of even order (hereafter double poles), corresponding to a finite truncation of the operator algebra as detailed in Section \ref{section:even}.
On the other hand, for an odd denominator $\rho$, we see both the double and simple poles.
The simple poles are responsible for the unitarity violation at each level, and the double ones lead to the logarithmic 4-pt functions. This is outlined in Section \ref{section:odd}.

\subsection{Hierarchical geodesics on the Poincar\'{e} disk: the Farey paths and the finite-$p$ amplitudes}
\label{section:poincare}
\begin{figure}
\begin{center}
\includegraphics[width=16cm]{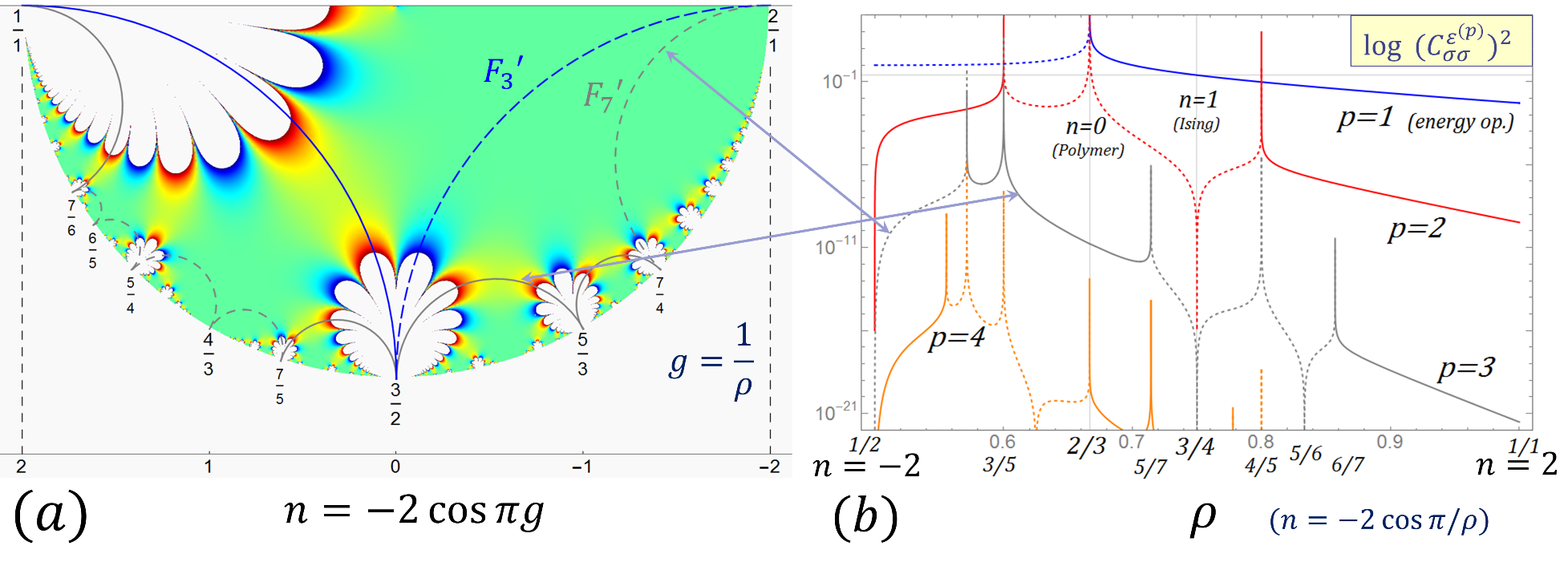}
\end{center}
\vspace{-0.8cm}
\caption{\label{figure:poincare}
(a) Poincar\'{e} disk and the Farey paths, $F'_3$ (blue) and $F'_7$ (gray). 
Each rational angle $g$ indicated by a fraction corresponds to a special CFT.
A portion of the disk is colored red (blue) if $\text{Re}\, \xi(q)$ is positive (negative), or colored green (light-gray) if the absolute value is almost infinitesimal (infinite).  
(b) The amplitudes in log-scale for $p=1$ (blue), $p=2$ (red), $p=3$ (gray), and $p=4$ (orange)  as a function of the parameter $\rho$.
The positive (negative) amplitude is represented by the solid (dashed) curve.
The up-(down-)ward spikes are poles (zeros).
The two-sided gray arrows indicate the possible map between the path $F'_7$ and the $p=3$ amplitude.
A crudest, yet intuitive, way for transforming the raw plot (b) into the geometrical picture (a) is rotating the former by $180^\circ$ and flipping the numerators/denominators of $\rho$ to obtain $g$.
}
\end{figure}

Let us now describe some possible analogy between our problem and certain geodesic paths in the hyperbolic geometry.
One could motivate this by following physicists' habit of taking some extremal case to identify the symmetry of the problem. 
Let us imagine $p\to \infty$ limit, where we anticipate the spikes everywhere, at every rational points of $\rho$.
Apparently, such a function might look peculiar, or even pathological. In retrospect, however, such an accumulation of the dense singularities would be rather commonly found as a section of the automorphic function near the infinity of the hyperbolic space.
An automorphic function is a function which is invariant under the action of some discrete groups such as 
$SL(2, \mathbb{Z})$. 
It is essentially defined on the single fundamental tile, and this tile is tessellated by iterating certain reflections.
As in the Escher's famous ``Circle Limit", these tiles are accumulated around the infinity.

Here the real-axis interval $|n|\leqslant 2$ is projected back to the lower boundary circle of the Poincar\'{e} disk as in Figure \ref{figure:poincare}-(a), and we measure its angle in the unit of $\pi$ by $g$. This angle is nothing but the Coulomb gas coupling \cite{difrancesco}, which takes the value $1<g<2$ and 
$g=\rho^{-1}$. 
In order to regularize the infinite-$p$ amplitude to get a finite-$p$ amplitude, it is necessary to escape from the singularities by entering inside the disk. 
To this end, the nome $q$ is introduced by 
\begin{align}
q=e^{i\pi g}=\frac{1}{2}\left(n-i\sqrt{4-n^2} \right),
\label{qnome}
\end{align}
which is the same as the parameter in the quantum group $U_q(sl_2)$ \cite{pasquier}. 
Although the $g$-$n$ projection relation in \eqref{qnome} is equivalent to the renowned one in \eqref{formulaBeffara},
to our knowledge, the direct embedding construction such as the one in Figure \ref{figure:poincare}-(a) has not been discussed.

By virtue of the clear distinction between the spectrum at the even/odd denominator $\rho$ observed in 
Section \ref{section:stern}, among the many possible automorphic functions, the elliptic modular function \cite{simon} $\lambda(q)$ may be useful.
Further, the $\xi(q)$ function is introduced by 
\begin{align}
\xi(q)=2-(\lambda(q)+\lambda(q)^{-1}),\quad
\lambda(q) = \left(\frac{\theta_2(q)}{\theta_3(q)}\right)^4=16q-128q^2+704q^3-3072q^4+\cdots.
\label{xi}
\end{align}
Both function has the following invariance under the transformations of the angle $g$, 
\begin{align}
\lambda(g)=\lambda(g+2),\qquad \lambda(g)=\lambda\left( \frac{g}{2g+1} \right)
\end{align}
corresponding to the two generators 
$\begin{pmatrix}
1 & 2 \\
0 & 1
\end{pmatrix}$
and
$\begin{pmatrix}
1 & 0 \\
2 & 1
\end{pmatrix}$
defining a congruence subgroup of the $SL(2, \mathbb{Z})$ denoted by $\Gamma(2)$.
Along the edges of the tiles generated by $\Gamma(2)$, the imaginary part of $\xi(q)$ vanishes as
$\text{Im}\ \xi(q)=\text{Im}\ \lambda(q)=0$. 
At the corners of the fundamental triangle tile, $\lambda(q)$ and $\xi(q)$ respectively take the set of values $\{0,1,\infty\}$ and $\{\infty, 0, \infty\}$, where the sign of $\infty$ depends on the incoming direction to the corner.
The choice of $\xi(q)$ is such that it has a double zero (up to reparameterization) at $\rho$ with an even denominator 
instead of any singularities, but is still ad-hoc. Accordingly, the following correspondence 
between the path and the amplitude 
may have minor exceptions in an idealized rule.  

To connect the zeros and the poles of the amplitude \eqref{CpssE} for finite-$p$, we introduce the Farey path in the following.
In the idealized rule, it is simply a path that connects the neighboring fractions of $g$ by the unique geodesic semicircle.
Each broccoli-like structure in Figure \ref{figure:poincare}-(a)
happens at the angle where infinitely many tiles share the corners, and
represents an essential singularity of $\xi(q)$. 
Such an existence of the essential singularity at an odd denominator $\rho$ makes 
the sign along the path diverse and interesting. 
For instance, if a Farey path enters the broccoli from a red region and escapes to a blue region,
we interpret it that the corresponding amplitude has a chance to cross a simple pole at that particular angle
since $\xi(q)$ changes its sign from $+\infty$ to $-\infty$. 
Similarly, if the path enters it and escapes from it passing the same color, the amplitude has a chance to hit a double pole, or just remains finite and the same sign. For the latter case,
one need to regularize the (idealized geodesic) path within the broccoli region to avoid the singularity. 

The first non-trivial examples of the Farey paths visit the angles in the following order:
\begin{align}
p=1\longrightarrow F_3:& ~~ \frac{1}{1},\, \frac{\hat 3}{2},\, \frac{2}{1}\\
p=2\longrightarrow F_5:& ~~ \frac{1}{1},\, \frac{\hat 5}{4},\, \frac{4}{3},\, \frac{3}{2},\, \frac{\hat 5}{3},\, \frac{2}{1}\\
p=3\longrightarrow F_7:& ~~ \frac{1}{1},\, \frac{\hat 7}{6},\, \frac{6}{5},\, \frac{5}{4},\, \frac{4}{3},\,\frac{\hat 7}{5},\, \frac{3}{2},\, \frac{5}{3},\, \frac{\hat 7}{4},\, \frac{2}{1},
\end{align}
where the hats are on the highest odd numerators for a given $p$, 
which corresponds to the simple poles of the amplitude.

The simplest case $F_3$ ($p=1$) corresponds to the amplitude of the energy operator.
The amplitude starts from $g=1/1$ ($n=2$, $c=1$) with a positive value, changes its sign at the unique simple pole at $g=3/2$ ($n=0$), and 
terminates at  $g=2/1$ ($n=-2$, $c=-2$) with a negative value. 
Similarly, along $F_3$, $\xi(q) \in \mathbb{R}$  changes its sign from $+\infty$ to $-\infty$ at $g=3/2$.
As $g=3/2$ is an essential singularity, the behavior of $\xi(q)$ along $F_3$ is the same as the amplitude for $p=1$ up to a monotonic reparameterization along the path to make it like a simple pole.
 
The Farey path $F_7$ is also shown (gray) as an example at the higher level, which amounts to the depth $p=3$ (gray frame) in Figure \ref{figure:Farey}. The amplitude for $p=3$ has 
three simple poles and three zeros. 
In this example, the amplitude  and the $\xi(q)$ function along the Farey path $F_7$ share 
exactly the same sign and the same location of the double zeros.



\section{Even and odd classes of the special points in the $O(n)$ CFT}\label{evenodd}
\subsection{Even denominator $\rho$: an algebra embedding to the minimal model} \label{section:even}
Let us now start our case study of even integer denominators. This is very simple, but important. 
It is well-known that the Kac scaling dimension $\Delta_{r,s}$ is invariant under the reflection of the $(r, s)$-plane with respect to the line $s=(m/m')r$ \cite{belavin}.  This also leads to a secret symmetry 
\begin{align}
\Delta_{r, s}\Longleftrightarrow \Delta_{\frac{m}{2}-r, \frac{m'}{2}-s}
\label{secret}
\end{align}
in addition to the standard identification 
$\Delta_{r, s}\Leftrightarrow \Delta_{m-r, m'-s}$ in the minimal model \cite{belavin}.
Now let $\rho=m/m'$ be an irreducible fraction with a denominator $m'$ even integer. 
This assumption yields a pair of integers 
$(\tilde{r}, \tilde{s})=\left((m-1)/2, m'/2 \right) \in \mathbb{N}^2$.
The spin field at $(r,s)=(1/2, 0)$ then becomes a degenerate field $(\tilde{r}, \tilde{s})$  
and the operator algebra closes.
In other words, one can embed the operator algebra
 into that of a certain minimal model $M_{m,m'}$ at these special cases of $\rho$ if $m'$ is an even integer
\footnote{
For {\it non-fundamental} fields,
this does not preclude the existence of the logarithmic 4-point functions at even $m'$.
A good example of the latter is found \cite{gori} for
the {\it boundary} operators $\sigma_{b}=\phi_{1,3}\neq \phi_{2,1}$ 
representing the connectivities of the Fortuin-Kasteleyn clusters in the Ising model ($m'=4$).
}.
According to \eqref{formulaBeffara}, for instance, the $O(n)$ CFT for $n=1$ is at $\rho=3/4$, and has the spin field at $(r,s)=(1,2)$: 
\begin{align}
\sigma=\phi_{\frac{1}{2},0} \Longleftrightarrow  \phi_{1,2} \qquad    \text{at}~~   \rho=3/4 ~~(n=1).  
\end{align}
In this case, the standard fusion rule $\phi_{1,2}\times \phi_{1,2}\sim \phi_{1,1}+\phi_{1,3}$ corresponds to,
\begin{align}
\sigma\times \sigma = \varepsilon^{(0)} + \varepsilon^{(1)}.
\label{isingOPE}
\end{align}
The infinite operator algebra \eqref{sigmasigma} for the $O(n)$ CFT truncates at $p=1$. 
Here one has only
one OPE coefficient to be determined. In passing, the formula \eqref{CpssE} naturally reproduces the famous result $\left[C^{\varepsilon}_{\sigma\sigma} \right]^2=\frac{1}{4}$ in the 2d Ising model as it should.
It is easy, but important, to show that each of the zeros for $p\geqslant 2$,
\begin{align}
0=\left[C^{\varepsilon^{(2)}}_{\sigma\sigma} \right]^2
=\cdots
=\left[C^{\varepsilon^{(\infty)}}_{\sigma\sigma} \right]^2
\label{truncate}
\end{align}
is not a simple zero, but the degree is even. Thus, all the amplitude in \eqref{CpssE} do not change their sign at $n=1$.
In summary, these infinitely many zeros realize a finite algebra at $\rho$ with an even denominator while they do not trigger the unitarity violation as $n$ crosses the corresponding value $n=-2\cos \pi m/m'$. 




\subsection{Odd denominator $\rho$: the logarithmic $4$-point function $\langle\sigma\sigma\sigma\sigma\rangle_{\text{phys}}$}\label{section:odd}
The odd denominator case is much more interesting. 
Consider an example the CFT at $\rho=4/5+\delta$, for which one has $n=\sqrt{2}$ as $\delta\to 0$.
One may observe the resonance between different channels due to
a pair of the scaling dimensions that differ by some integers.
This can be easily seen by organizing the right hand side of \eqref{sigmasigma} by using the field identification $\phi_{r, s}= \phi_{m-r, m'-s}$ which holds exactly at $\delta=0$:
\begin{align}
\sigma \times \sigma =& \phi_{1,1} +\phi_{1,3} + \phi_{1,5} + \phi_{1,~7}\ + \phi_{1,~9}\, +\cdots\\
                                   \overset{\delta=0}{=}& \phi_{3,4} +\phi_{3,2} +  \uwave{\phi_{3,0}} + \phi_{3,-2}+ \phi_{3,-4}+ \cdots \label{pivot}
\end{align}
By the well-known relation $\Delta_{r,-s}=\Delta_{r,s} + rs$, one has $\Delta_{1,9}=\Delta_{1,1}+12$, $\Delta_{1,7}=\Delta_{1,3}+6$, thereby the claimed resonances happen.
For instance, a primary contribution of the 5-th channel ($\varepsilon^{(4)}=\phi_{1,9}$) starts to mix with a level-12 descendant contribution of the 1-st channel ($\varepsilon^{(0)}=\phi_{1,1}$; the identity operator) as shown in \eqref{resonance12}. 
A ``primary" contribution here means that it comes from the leading contibution in the 
block $|F_{\Delta_{1,9}}(z)|^2 \sim |z|^{-2\Delta_{1,9}}$. 
Similarly, the 2-nd ($\varepsilon^{(1)}=\phi_{1,3}$; the energy operator) and 4-th channel ($\varepsilon^{(3)}=\phi_{1,7}$) mix with each other at level 6 as in \eqref{resonance6}.
These resonances are summarized as follows.
\begin{align}
\langle \sigma\sigma\sigma\sigma \rangle &= 
(C_{\sigma\sigma}^{1,1})^2 |F_{\Delta_{1,1}}(z)|^2&\Longleftarrow\ &\mathcal{O}(1)=
\left(C_{\sigma \sigma }^{1,1}\right)^2
,~    \Delta_{1,9}=\Delta_{1,1}+12+\mathcal{O}(\delta),
&&p=0 ~(\spadesuit)  \label{resonance12}\\ 
&+ (C_{\sigma\sigma}^{1,3})^2 |F_{\Delta_{1,3}}(z)|^2&\Longleftarrow\ &\mathcal{O}(1)
=\left(C_{\sigma \sigma }^{1,3}\right)^2,
~     \Delta_{1,7}=\Delta_{1,3}+6+\mathcal{O}(\delta),   &&p=1 ~(\clubsuit) \label{resonance6}\\
&+ { (C_{\sigma\sigma}^{1,5})^2 |F_{\Delta_{1,5}}(z)|^2}&\Longleftarrow\ & 
\mathcal{O}(\delta^{-1})=(C_{\sigma\sigma}^{1,5})^2, &&p=2 ~(\diamondsuit)   \\
&+  (C_{\sigma\sigma}^{1,7})^2 |F_{\Delta_{1,7}}(z)|^2&\Longleftarrow\ &  -(A_1/\delta)^2 +\mathcal{O}(\delta^{-1})
=\left(C_{\sigma \sigma }^{1,7}/C_{\sigma \sigma }^{1,3}\right)^2 ,   &&p=3 ~(\clubsuit)\label{p3}\\
&+ (C_{\sigma\sigma}^{1,9})^2 |F_{\Delta_{1,9}}(z)|^2&\Longleftarrow\ & 
-(A_2/\delta)^2 +\mathcal{O}(\delta^{-1})=\left(C_{\sigma \sigma }^{1,9}/C_{\sigma \sigma }^{1,1}\right)^2,  &&p=4 ~(\spadesuit)\label{p4}\\
&+\cdots, &&\nonumber
\end{align}
where the simplified notation $C_{\sigma \sigma }^{r,s}=C_{\sigma \sigma }^{\phi_{r,s}}$ is used.
This shows the first resonance $(\spadesuit)$ occurs between 
$\varepsilon^{(0)} \leftrightarrow \varepsilon^{(4)}$ 
and the second $(\clubsuit)$ occurs between $\varepsilon^{(1)} \leftrightarrow \varepsilon^{(3)}$. 
Note that this phenomenon is due to the exisitence of the pivot operator $p=2$ $(\diamondsuit)$  
at $\delta=0$ as indicated by the wavy line in \eqref{pivot}, where the duality $\phi_{1,5}\Leftrightarrow \phi_{3,0}\ $ holds.
 This operator comes as a simple pole $\mathcal{O}(\delta^{-1})$.

The amplitude ratios between these resonating channels (the ratio of the squared OPE coefficients) may be computed from \eqref{CpssE}, resulting in the leading double poles in the deviation parameter $\delta$ as in \eqref{p3} and \eqref{p4}.
The coefficients $A_1$ and $A_2$ may be obtained as rather large products of the gamma functions, which 
could look non-instructive at first: 
\begin{align}
A_1=&
\scalebox{0.85}{$\displaystyle
\frac{1}{60}\sqrt{\frac{-\Gamma \left(-\frac{19}{5}\right) \Gamma \left(-\frac{18}{5}\right) \Gamma \left(-\frac{14}{5}\right) \Gamma \left(-\frac{11}{5}\right) \Gamma \left(-\frac{6}{5}\right) \Gamma \left(\frac{11}{10}\right)^2 \Gamma \left(\frac{8}{5}\right)^4 \Gamma \left(\frac{19}{10}\right)^2 \Gamma\left(\frac{21}{10}\right)^2 \Gamma \left(\frac{12}{5}\right)^3 \Gamma \left(\frac{29}{10}\right)^2}{\Gamma \left(-\frac{19}{10}\right)^2 \Gamma \left(-\frac{7}{5}\right)^3 \Gamma \left(-\frac{11}{10}\right)^2 \Gamma \left(-\frac{9}{10}\right)^2 \Gamma \left(-\frac{3}{5}\right)^4 \Gamma \left(-\frac{1}{10}\right)^2 \Gamma \left(\frac{11}{5}\right) \Gamma \left(\frac{16}{5}\right) \Gamma \left(\frac{19}{5}\right) \Gamma \left(\frac{23}{5}\right) \Gamma \left(\frac{24}{5}\right)}} $}
\nonumber\\
=&\frac{3^2\cdot 7^1\cdot 11^1\cdot 19^1}{2^{23}\cdot 5^7\cdot 13^1},
\label{A1}
\\
A_2=&\scalebox{0.85}{$\displaystyle
\frac{1}{60}\sqrt{\frac{\Gamma \left(-\frac{27}{5}\right) \Gamma \left(-\frac{26}{5}\right) \Gamma \left(-\frac{23}{5}\right) \Gamma \left(-\frac{22}{5}\right) \Gamma \left(-\frac{19}{5}\right) \Gamma \left(-\frac{18}{5}\right) \Gamma \left(-\frac{14}{5}\right) \Gamma \left(-\frac{6}{5}\right) \Gamma \left(-\frac{2}{5}\right) }
{  \Gamma \left(-\frac{27}{10}\right)^2 \Gamma \left(-\frac{11}{5}\right)^3 \Gamma \left(-\frac{19}{10}\right)^2 
\Gamma \left(-\frac{17}{10}\right)^2 \Gamma
   \left(-\frac{7}{5}\right)^3 \Gamma \left(-\frac{11}{10}\right)^2 \Gamma \left(-\frac{9}{10}\right)^2 \Gamma \left(-\frac{3}{5}\right)^3 \Gamma \left(-\frac{3}{10}\right)^2 \Gamma \left(-\frac{1}{10}\right)^2 }}
$}\nonumber\\
&\cdot\scalebox{0.85}{$\displaystyle\sqrt{\frac{\Gamma\left(\frac{3}{10}\right)^2 \Gamma \left(\frac{2}{5}\right) \Gamma \left(\frac{4}{5}\right)^3 \Gamma \left(\frac{11}{10}\right)^2 \Gamma \left(\frac{13}{10}\right)^2 \Gamma \left(\frac{8}{5}\right)^3 \Gamma \left(\frac{19}{10}\right)^2 \Gamma \left(\frac{21}{10}\right)^2 \Gamma \left(\frac{12}{5}\right)^3 \Gamma   \left(\frac{27}{10}\right)^2 \Gamma \left(\frac{29}{10}\right)^2 \Gamma \left(\frac{16}{5}\right)^3 \Gamma \left(\frac{37}{10}\right)^2}
{\Gamma \left(\frac{1}{5}\right)^3 \Gamma \left(\frac{3}{5}\right) \Gamma \left(\frac{7}{10}\right)^2 \Gamma   \left(\frac{7}{5}\right) \Gamma \left(\frac{11}{5}\right) \Gamma \left(\frac{19}{5}\right) \Gamma \left(\frac{23}{5}\right) \Gamma \left(\frac{24}{5}\right) \Gamma \left(\frac{27}{5}\right) \Gamma \left(\frac{28}{5}\right) \Gamma \left(\frac{31}{5}\right) \Gamma \left(\frac{32}{5}\right)}}$}
\nonumber \\
=&\frac{3^3\cdot 7^2\cdot 11^2\cdot 17^2\cdot 19^1}{2^{50}\cdot 5^7\cdot 13^3\cdot 23^1}.
\label{A2}
\end{align}
Both these coefficients, however, suggestively reduce to nice fractions  
as above after some repeated uses of the elementary relation $\Gamma(x)\Gamma(1-x)=\pi/\sin \pi x$.
Such reduction can not be a coincidence.

In fact, we can show that the double pole contributions at higher operators are exactly canceled out by the descendant contributions of the lower operators.
The key is the Zamolodchikov recursion relation \cite{zamolodchikov}. If we focus on the dependence on the scaling dimension $\Delta$ of the intermediate conformal blocks, then they have simple poles at the degenerate dimensions:
\begin{align}
F_{\Delta}=g_{\Delta} + \sum_{(r,s)\in \mathbb{N}^2} \frac{R_{r, s}z^{rs}}{\Delta -\Delta_{r,s}}F_{\Delta_{r,s}+rs},
\end{align}
where the residue is proportional to the conformal block with shifted $\Delta$ with the coefficient $R_{r,s}$ conjectured by Zamolodchikov, which is the following finite product of linear combinations of the external charges $\{\alpha_i\}$
and the screening charges $\alpha_\pm$:
\begin{align}
R_{r,s}=B_{r,s}\cdot P_{r,s}(c, \{\alpha_i\}),\qquad &B_{r,s}=-\frac{1}{2}\prod_{j, k}\lambda^{-1}_{j,k} ~~~(|j|\leqslant r,~ |k|\leqslant s,~ (j,k)\neq (0,0), (r,s)), \\
&P_{r,s}(c, \{\alpha_i\})=\prod_{\pm, p, q}
\left(\alpha_1\pm \alpha_2 +\frac{\lambda_{p,q}}{2}\right)
\left(\alpha_3\pm \alpha_4 +\frac{\lambda_{p,q}}{2}\right),
\end{align}
where $(p, q)$ runs $p=-r+1, -r+3, \cdots, r-1$ and $q=-s+1, -s+3, \cdots, s-1$ and 
$\lambda_{p,q}=p\alpha_{+} + q\alpha_{-}$.
Importantly, our assumptions of the denominator $m'$ be odd, and of  the external dimensions be all non-degenerate ($\alpha_i=\alpha_{1/2, 0}$ for $i=1,2,3,4$) guarantee non-vanishing of the residue coefficients $R_{r,s}$ for certain pairs $(r,s)$, such as $(r,s)=(3,4)$ and $(r,s)=(3,2)$, which respectively correspond to the resonating channels in \eqref{resonance12} and \eqref{resonance6}. 
And this property makes $F_{\Delta_{1,1}}$ and $F_{\Delta_{1,3}}$ the {\it singular} conformal blocks 
\cite{viti} as $\delta\to 0$:  
\begin{align}
F_{\Delta_{1,1}} \supset \frac{R_{3,4}\ z^{12+\mathcal{O}(\delta)}}{\Delta_{1,1}-\Delta_{3,4}}=\frac{A_2 z^{12}}{\delta}
\left( 1+\mathcal{O}(\delta)\ln z \right),\,
F_{\Delta_{1,3}}\supset 
\frac{R_{3,2}\ z^{6+\mathcal{O}(\delta)}}{\Delta_{1,3}-\Delta_{3,2}}=\frac{A_1 z^{6}}{\delta}
\left( 1+\mathcal{O}(\delta)\ln z \right).
\end{align}
Then it is possible to show that the left hand side becomes simple pole whose coefficients, when squared, are exactly the same as \eqref{A1} and \eqref{A2} from
 our integral formula computation except the signs. 
Since the charge at a rational $\rho$ is at most the quadratic irrational, 
the gamma functions can not remain in \eqref{A1} and \eqref{A2}, 
which rationalizes the nice reduction we mentioned above. 
Due to the cancellation at the leading $\delta^{-2}$, one should expand the resonant powers of $z$ in
\eqref{resonance12} and \eqref{resonance6}, which results in the logarithmic 4-point function. 
Since all the dominant terms contribute as $\delta^{-1}$, the physical correlation 
function 
is actually proportional to 
\footnote{A similar observation has been made in the percolation CFT ($c=0$) \cite{dotsenko},
where one may consider the combination $(q-1) \langle \sigma\sigma\sigma\sigma \rangle_{\text{CFT}}$
in the $q\to 1$ limit of the $q$-state Potts model. 
Our method also turns out to be efficient in computing such correlation functions.}
$\delta \langle \sigma\sigma\sigma\sigma \rangle_{\text{CFT}}$.
As an important example, at the polymer CFT $(n=0, c=0)$, one has $\rho=2/3+\delta$ and $\delta=\mathcal{O}(n)$.
Then the observable 4-point function is
 $\langle \sigma\sigma\sigma\sigma \rangle_{\text{phys}}=n \langle \sigma\sigma\sigma\sigma \rangle_{\text{CFT}}$
in $n\to 0$.
Besides the application to the polymer, this gives us a clue to the way we should deal with the simple pole at $n=0$ of the energy operator amplitude, and with all its generalization at $\rho$ with the odd denominator.

\section{Conclusion}
The 4-point function $\langle\sigma\sigma\sigma\sigma\rangle$ of the $O(n)$ fundamental field
plays a central role in the conformal bootstrap in general dimensions.
In contrast with 4-point functions that contain degenerate fields \cite{belavin, dotsenkofateev}, 
this 4-point function $\langle\sigma\sigma\sigma\sigma\rangle$ remains elusive even in $d=2$ for generic values of $n = -2\cos \pi/\rho$ since all the fields in the correlation function are non-degenerate.  
We have seen, however, if the standard identification \cite{difrancesco} $\varepsilon\to \phi_{1,3}$ and $\sigma \to \phi_{1/2,0}$ is assumed, one can compute the important amplitudes \eqref{CpssE} that correspond to the $O(n)$ singlet contribution. Seeing the unitarity from a higher perspective, 
it would be important to quantify to what degree such amplitudes can be negative. 
By brushing up the knowledge obtained here, it is possible to discuss that the positivity violation is very weak at $n=0.99$, for instance, like it is weak at $d=3.99$ \cite{hogervorst}. On the practical side, 
such direction could be a clue for approaching the real non-unitary CFT from the (almost unique) unitary solution to the crossing symmetry obtained in non-integer $d$ \cite{vichi} and in non-integer $n$ \cite{shimada}.
We have seen that the sign-change occurs only at the simple poles, 
whose exact locations are encoded in the behavior of the automorphic $\xi(q)$-function \eqref{xi} along the Farey path.
Although the correspondence between these ideal Farey paths on the Poincar\'{e} disk and the singlet comformal amplitudes beyond the simple poles is not perfect as it is, 
it would be interesting to make it precise or to see if such hierarchical structure is ubiquitous in the CFT with infinitely many primary operators.


\ack
This work is supported by KAKENHI  16K05491.

\end{document}